\documentclass[prl, reprint, superscriptaddress]{revtex4-1}

\usepackage{amssymb}
\usepackage{amsmath}
\usepackage{graphicx}
    \graphicspath{{figures/}}

\begin{document}

\title{Extreme THz fields from two-color filamentation of mid-infrared laser
       pulses}

\author{Vladimir~Yu.~Fedorov}
\affiliation{Science Program, Texas A\&M University at Qatar, P.O. Box 23874,
             Doha, Qatar}
\affiliation{P.~N.~Lebedev Physical Institute of the Russian Academy of
             Sciences, 53 Leninskiy Prospekt, 119991, Moscow, Russia}
\email{v.y.fedorov@gmail.com}

\author{Stelios~Tzortzakis}
\affiliation{Science Program, Texas A\&M University at Qatar, P.O. Box 23874,
	Doha, Qatar}
\affiliation{Institute of Electronic Structure and Laser (IESL), Foundation for
             Research and Technology - Hellas (FORTH), P.O. Box 1527, GR-71110
             Heraklion, Greece}
\affiliation{Materials Science and Technology Department, University of Crete,
             71003, Heraklion, Greece}
\email{stzortz@iesl.forth.gr}

\date{\today}

\begin{abstract}
    Nonlinear THz photonics is probably the last frontier of nonlinear optics.
    The strength of both the electric and the magnetic fields of these
    ultrashort low frequency light bunches opens the way to exciting science and
    applications.
    Progress in the field though is slow because of the deficiency in suitable
    sources.
    Here we show that two-color filamentation of mid-infrared 3.9\,$\mu$m laser
    pulses allows one to generate single cycle THz pulses with multi-millijoule
    energies and extreme conversion efficiencies.
    Moreover, the focused THz peak electric and magnetic fields reach values of
    GV/cm and kT, respectively, exceeding by far any available quasi-DC field
    source today.
    These fields enable extreme field science, including into other,
    relativistic phenomena.
    Besides, we elucidate the origin of this high efficiency, which is made up
    of several factors, including a novel mechanism where the harmonics produced
    by the mid-infrared pulses strongly contribute to the field symmetry
    breaking and enhance the THz generation.
\end{abstract}

\maketitle

{\it Introduction.}---%
The terahertz (THz) frequency range (0.1-10\,THz) is a part of the
electromagnetic spectrum located at the junction between the microwave and
optical frequencies.
For many reasons THz frequencies attract a lot of interest in recent
years~\cite{Zhang2017,Tonouchi2007}.
Since structural absorption resonances of many molecules belong to the THz
frequency band, THz spectroscopy becomes a unique tool for matter studies.
Moreover, THz frequencies lie at the boundary of frequency ranges that are
characteristic for high-frequency electronics and photonics.
Therefore, THz devices are expected to be a connecting link between these
technologies.
In addition, THz radiation penetrates through a variety of non-conducting
materials like clothing, paper, wood, masonry, plastic and ceramics.
But unlike X-rays, THz radiation is not an ionizing radiation and does not
damage test materials, which opens great opportunities for its application in
industrial quality control, homeland security, or medical diagnostics and
treatment.

Although the THz frequency range has very rich scientific and technological
potential, it remains underexplored due to lack of intense THz sources and
sensitive THz detectors.
The progressive appearance of suitable bright THz sources opens a new era for
studies of extreme THz field-matter interactions, nonlinear THz spectroscopy and
imaging.
To date, there are two major techniques for the generation of intense THz pulses
on tabletop setups~\cite{Lewis2014,Reimann2007}: optical rectification in
nonlinear crystals~\cite{Huang2014,Vicario2014,Shalaby2015} and two-color
filamentation (photoionization of gases by dual-frequency laser
fields)~\cite{Kim2008,Oh2014,Kuk2016}.
Optical rectification has THz conversion efficiencies (ratio of generated THz
energy to energy of input laser pulse) that can reach 3.7\%~\cite{Huang2014}.
Nevertheless, the spectral bandwidth of these sources is limited to frequencies
below 5\,THz and the generated THz pulses energies are also limited, because of
the damage threshold of the crystals, with the highest energy reported to date
reaching 0.9\,mJ~\cite{Vicario2014}.
In turn, two-color filamentation presents lower THz conversion efficiencies
($\sim$0.01\%) and produces less intense THz pulses with energies up to
30\,$\mu$J~\cite{Kim2008,Oh2014,Kuk2016}.
However, THz pulses generated by two-color filamentation have much larger
spectral bandwidths ($>50$\,THz) and can be generated at remote
distances~\cite{Wang2010,Wang2011,Daigle2012}, which allows to overcome a number
of THz propagation issues, such as high absorption in atmospheric water vapor
and diffraction.

Up to date, most experiments on two-color filamentation were conducted using
Titanium:Sapphire (Ti:Sa) laser sources with central wavelength around
0.8\,$\mu$m and their second harmonic.
A study using longer wavelengths, in the near infrared, showed an enhancement of
the THz conversion efficiency with increasing pump wavelength but only up to
1.8\,$\mu$m, while beyond this point the efficiency dropped
again~\cite{Clerici2013}.
Also, a theoretical study showed stronger THz generation compared to 0.8\,$\mu$m
using single color mid-infrared pulses with Particle in Cell (PIC) simulations,
not considering though nonlinear propagation effects~\cite{Wang2011OL}.
Recently though, filamentation of mid-infrared (mid-IR) laser pulses with
3.9\,$\mu$m central wavelength was for the first time demonstrated
experimentally~\cite{Mitrofanov2015}.
The subsequent experimental and theoretical studies revealed that compared to
0.8\,$\mu$m laser pulses, filaments produced by 3.9\,$\mu$m are longer with
wider plasma channels and the generated supercontinuum is extremely broad
ranging up to harmonics of 15th order~\cite{Panagiotopoulos2015,Mitrofanov2015,
Mitrofanov2016a,Mitrofanov2016b,Panagiotopoulos2016,Panov2016}.

In this work we study numerically the generation of THz radiation by two-color
filamentation of 3.9\,$\mu$m laser pulses.
We show that compared to 0.8\,$\mu$m pulses, the THz conversion efficiency in
mid-IR filaments is two orders of magnitude higher, the energy of the generated
THz pulses reaches the multi-millijoule level, and the strength of the THz
fields can reach the GV/cm range.
Thus, two-color filamentation of mid-IR laser pulses emerges as the ultimate
source for extreme THz science since it allows one to generate THz radiation
with unprecedented efficiency and extremely high energy that largely overcomes
all other approaches while energy scaling does not suffer from undesired effects
like damage of crystals in the optical rectification.

{\it Model.}---%
To simulate the two-color filamentation of mid-IR laser pulses in air we use the
Unidirectional Pulse Propagation Equation (UPPE)~\cite{Kolesik2002,Kolesik2004}
coupled with the kinetic equation for plasma concentration~\cite{Couairon2011}
(see the Supplemental Material~\cite{Supplement}).
This model takes into account nonparaxial propagation of polychromatic fields
without any kind of envelope approximations.
It includes dispersion of all orders, cubic Kerr nonlinearity, defocusing in
plasma, inverse Bremsstrahlung, photoionization with corresponding energy
losses, and avalanche ionization.
We use a realistic dispersion model of dry atmospheric air (zero relative
humidity) that takes into account spectral lines of $N_2$, $O_2$, and $CO_2$
gases from the HITRAN database~\cite{HITRAN}.

Our initial condition for the UPPE equation is the following two-color field
$E$:
\begin{equation*}
    E = \exp\left(-\frac{r^2}{2a_0^2} - \frac{t^2}{2\tau_0^2}\right)
        \left[E_1 \cos\left(\omega_0t\right)
            + E_2 \cos\left(2\omega_0t\right)\right],
\end{equation*}
where $r^2=x^2+y^2$, $a_0=4/2\sqrt{\log{2}}$\,mm is the beam size (4\,mm FWHM),
$\tau_0=100/2\sqrt{\log{2}}$\,fs is the pulse duration (100\,fs FWHM),
$\omega_0$ is the central frequency, while $E_1$ and $E_2$ being the amplitudes
of the fundamental and second harmonic pulses, respectively.
The initial pulse is focused by a lens with a focal distance $f=200$\,mm.
To simulate the focusing we multiplied each Fourier harmonic of the field $E$ by
a factor $\exp\left[-i(\omega/c_0)r^2/(2f)\right]$, where $\omega$ is the
frequency of the corresponding harmonic and $c_0$ is the speed of light in
vacuum.
In our simulations the central wavelength $\lambda_0$ of the fundamental pulse
is equal to 3.9\,$\mu$m.
Also, in order to have a reference for comparison we simulated two-color
filamentation of the same laser pulse but with $\lambda_0=0.8$\,$\mu$m.
The energy $W$ of the initial two-color pulse is equal to 29\,mJ for
$\lambda_0$=3.9\,$\mu$m and 1.23\,mJ for $\lambda_0$=0.8\,$\mu$m.
For both cases the fundamental and second harmonic pulses hold, respectively,
95\% and 5\% of this energy.
The energy $W$ for each wavelength was chosen in such a way that the power of
the corresponding single-color pulse at wavelength $\lambda_0$ is equal, in both
cases, to 1.2$P_{cr}$, where $P_{cr}$ is the critical power of self-focusing in
air at this wavelength.

\begin{figure}[t] \centering
    \includegraphics{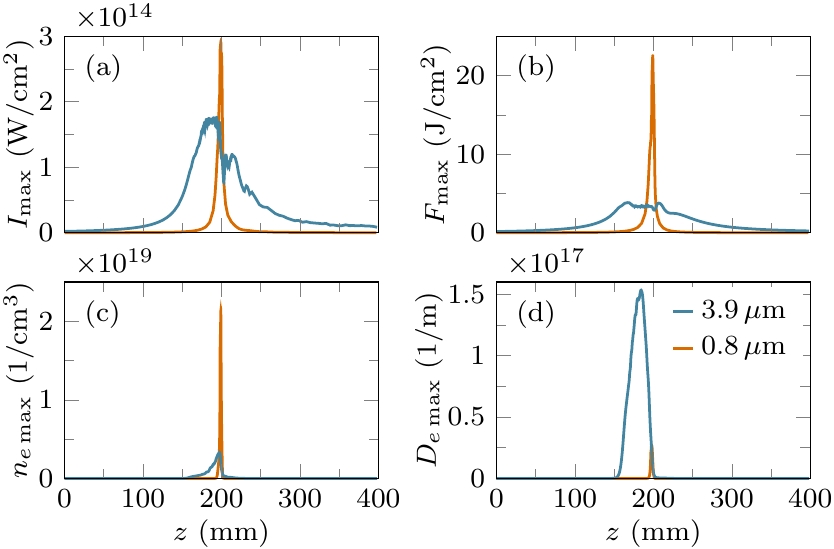}
    \caption{\label{fig:zplot}%
             (a) Peak intensity $I_\mathrm{max}$, (b) peak fluence
             $F_\mathrm{max}$, (c) peak plasma concentration
             $n_{e\,\mathrm{max}}$, and (d) plasma concentration integrated over
             radius, $D_{e\,\mathrm{max}}$, versus propagation distance $z$ for
             the 3.9\,$\mu$m and the 0.8\,$\mu$m two-color laser pulses.}
\end{figure}

{\it Results.}---%
Using the above model we numerically simulated the two-color filamentation of
3.9\,$\mu$m and 0.8\,$\mu$m pulses in air.
Figure~\ref{fig:zplot} shows the dependence of several filamentation integral
parameters on propagation distance for both wavelengths.
One can see that the peak intensity, peak fluence, and peak plasma concentration
in the case of 3.9\,$\mu$m pulse are several times lower than for the
0.8\,$\mu$m pulse.
However, the filament produced by the 3.9\,$\mu$m pulse is about 3 times longer.
Moreover, much higher integrated over radius plasma concentration shows that the
plasma channel produced by the 3.9\,$\mu$m pulse is considerably wider compared
to the 0.8\,$\mu$m pulse.

\begin{figure}[t] \centering
    \includegraphics{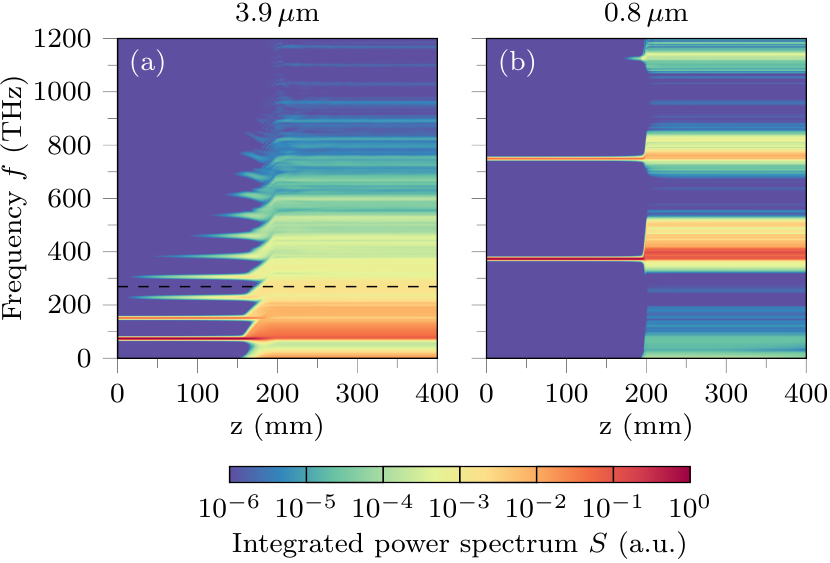}
    \caption{\label{fig:iSzf}%
             Dependence of integrated pulse power spectrum $S$ on propagation
             distance $z$ and frequency $f$ for 3.9\,$\mu$m (a) and 0.8\,$\mu$m
             (b) two-color laser pulses.
             The black dashed line separates the harmonics of 4th order and
             higher.}
\end{figure}

Figure~\ref{fig:iSzf} shows the dependence of the integrated power spectrum $S$
on propagation distance $z$ and frequency $f$ for the two wavelengths.
One can see that filamentation of 3.9\,$\mu$m two-color pulse is accompanied by
generation of extremely broad supercontinuum, which includes all even and odd
harmonics up to at least the 15th order.
Though the most intriguing result is the impressive energy transferred to the
THz part of the spectrum with the 3.9\,$\mu$m pulses compared to the case of the
0.8\,$\mu$m ones.
The THz conversion efficiency (for frequencies $f$ lying below 40\,THz) for
0.8\,$\mu$m pulses is 0.06\%, while the one for 3.9\,$\mu$m pulses is 6.7\%,
that is more than two orders of magnitude higher.
Note that this conversion efficiency is the highest reported to date compared to
any of the known approaches for THz generation.

\begin{figure}[t] \centering
    \includegraphics{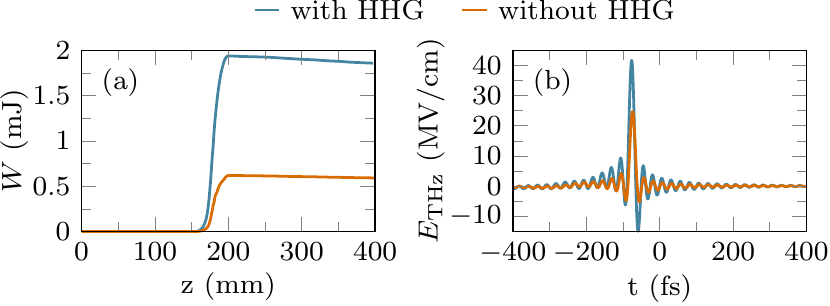}
    \caption{\label{fig:thz}%
             (a) Energy of THz pulse $W$ versus propagation distance $z$.
             (b) THz electric field versus time $t$ at $z=190$\,mm.
             THz pulses are generated during two-color filamentation of
             3.9\,$\mu$m pulses in the case of air with (cyan) or without
             (orange) high harmonics generation.}
\end{figure}

Figure~\ref{fig:thz}a shows how the energy of the THz pulse generated during
two-color filamentation of 3.9\,$\mu$m pulses depends on the propagation
distance $z$.
One can see that close to the end of the filament ($z\simeq220$\,mm) the energy
of the THz pulse reaches almost 2\,mJ.
Note that the decrease of the THz energy at longer propagation distances is
purely numerical and is due to losses of the diffracting THz beam in the
absorbing boundary layers located at the end of the numerical grid.
In turn, the peak THz energy generated by 0.8\,$\mu$m pulses is only about
0.8\,$\mu$J.
In other words, the THz pulse generated by the 3.9\,$\mu$m pulse is 2500 times
(three orders of magnitude!) more energetic than the one generated by
0.8\,$\mu$m pulses.

Figure~\ref{fig:thz}b shows the on-axis THz electric field generated by
3.9\,$\mu$m pulses at a distance $z=190$\,mm (in the middle of the filament).
One can see that the field strength (the amplitude from minimum to maximum of
the field) of the generated THz pulse reaches 56\,MV/cm, which exceeds the field
strengths obtained in the most efficient experiments with optical
rectification~\cite{Vicario2014}.

\begin{figure}[t] \centering
    \includegraphics{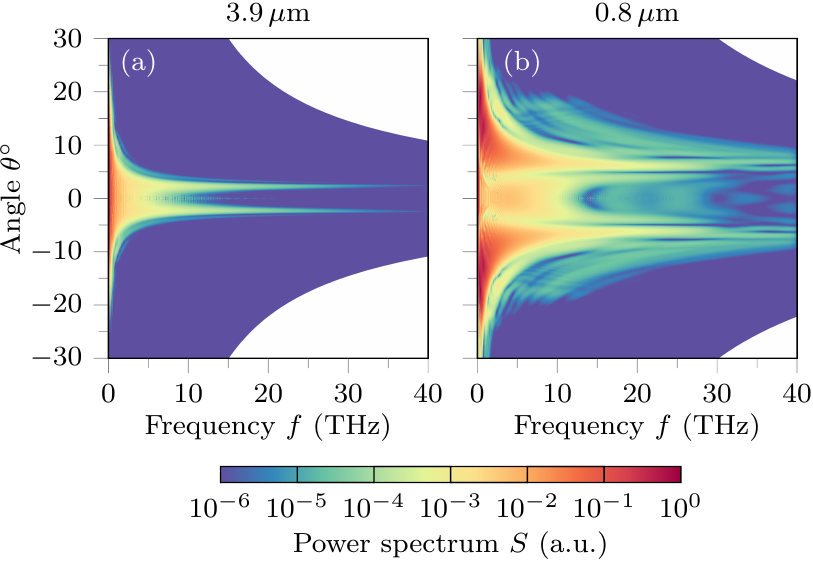}
    \caption{\label{fig:Sfangle}%
             Dependence of pulse power spectrum $S$ on frequency $f$ and angle
             $\theta$ for 3.9\,$\mu$m (a) and 0.8\,$\mu$m (b) two-color laser
             pulses at distance $z=220$\,mm.}
\end{figure}

Figure~\ref{fig:Sfangle} shows the angularly-resolved frequency spectrum of the
THz pulse generated during two-color filamentation of 3.9\,$\mu$m and
0.8\,$\mu$m pulses.
One can see that in both cases the THz radiation is emitted into a cone, where
higher frequencies propagate at smaller angles.
Also, we see that in average the angle of the THz emission cone is smaller in
the case of $\lambda_0=3.9$\,$\mu$m ($\sim7^\circ$ for 0.8\,$\mu$m and
$\sim2^\circ$ for 3.9\,$\mu$m pulses).
Thus, an extra advantage of the THz radiation produced during two-color
filamentation of 3.9\,$\mu$m is its better directionality compared to the case
of 0.8\,$\mu$m pulses.

{\it Discussion.}---%
As we have already seen two-color filamentation of mid-IR laser pulses allows
one to generate THz pulses of very high energy and field strength.
High efficiency of THz generation by mid-IR two-color pulses is made up of
several factors.
Maybe the most intriguing of them is related to the highly efficient generation
of high harmonics during mid-IR filamentation.
In Fig.~\ref{fig:iSzf} we see that the spectrum produced by 3.9\,$\mu$m pulses
consists of all harmonics, both even and odd, up to the 15th order.
Thus, filamentation of two-color 3.9\,$\mu$m pulses is accompanied by generation
of a multiple number of secondary dual frequency pulses ($2\omega$--$4\omega$,
$3\omega$--$6\omega$, $4\omega$--$8\omega$, etc.).
Each of these secondary pulses contribute to the field symmetry breaking and
support further THz generation.
In order to test this hypothesis, we repeated the simulations of two-color
filamentation with 3.9\,$\mu$m pulses, but during these simulations at each
propagation step we filtered out all harmonics of order 4 and higher (the black
dashed line in Fig.~\ref{fig:iSzf} shows the boundary of the spectral filter).
The energy and the electric field of the THz pulses obtained from these
simulations are plotted in Fig.~\ref{fig:thz}.
One can see that without higher harmonics the energy of the generated THz pulse
drops down by 3 times.

In addition, mid-IR laser pulses produce much stronger photocurrents compared to
0.8\,$\mu$m pulses.
This can be seen from the following estimations.
The velocity $v$ of a free electron under the action of the Lorentz force
produced by a monochromatic field of amplitude $A$ and frequency $\omega_0$ is
given by the equation $dv/dt=q_e/m_eA\cos(\omega_0t)$, where $q_e$ and $m_e$ are
the charge and mass of electron, respectively.
After integration we find $v=(q_e/m_e)(A/\omega_0)\sin(\omega_0t)$ (the initial
velocity of free electrons after ionization is assumed to be zero).
Therefore the average electron velocity $\sqrt{v^2}=(q_e/m_e)(A/2\omega_0)$ is
proportional to the wavelength $\lambda_0=2\pi c_0/w_0$.
Thus, the photocurrent $J=q_ev$ produced by 3.9\,$\mu$m pulses and, as a
consequence the generated THz field, is about five times stronger compared to
0.8\,$\mu$m pulses.

Another factor contributing to the highly efficient THz generation is a smaller
walk-off between the fundamental and the second harmonic.
According to our dispersion model (see the Supplemental
Material~\cite{Supplement}) the walk-off between 0.4\,$\mu$m and 0.8\,$\mu$m
pulses is $81$\,fs/m.
In turn the walk-off between $1.95$\,$\mu$m and $3.9$\,$\mu$m pulses is only
$1.26$\,fs/m for air with CO$_2$ and $3.04$\,fs/m for air without CO$_2$.
That is, the walk-off for $3.9$\,$\mu$m two-color pulses is at least 20 times
less than for 0.8\,$\mu$m pulses.
In turn, keeping in mind the Cherenkov mechanism of THz
generation~\cite{Johnson2013}, smaller walk-off between the THz and fundamental
pulses explains the better directionality of THz radiation in the case of mid-IR
pulses (see Fig.~\ref{fig:Sfangle}).

Besides, the THz generation efficiency is higher for $3.9$\,$\mu$m two-color
pulses because it produces $\sim3$ times longer plasma channels that contain
$\sim45$ times more free electrons (see Fig.~\ref{fig:zplot}).

It is interesting that in our studies we did not reveal any effect of the CO$_2$
gas on THz generation efficiency, as one could suggest because of the absorption
line near the 3.9\,$\mu$m and the corresponding anomalous dispersion (see the
Supplemental Material~\cite{Supplement}).
For air with and without CO$_2$ the results of our simulations are so close that
it is very hard to see any difference on any of the above figures.

To sum up, the much higher efficiency of THz generation by 3.9\,$\mu$m two-color
pulses compared to 0.8\,$\mu$m ones can be explained by the extra field symmetry
breaking due to higher harmonics, the 5 times stronger photocurrents, the tens
of times smaller walk-off between the fundamental and its second harmonic, the 3
times longer plasma channels, and the 45 times more total free electrons.

\begin{figure}[t] \centering
    \includegraphics{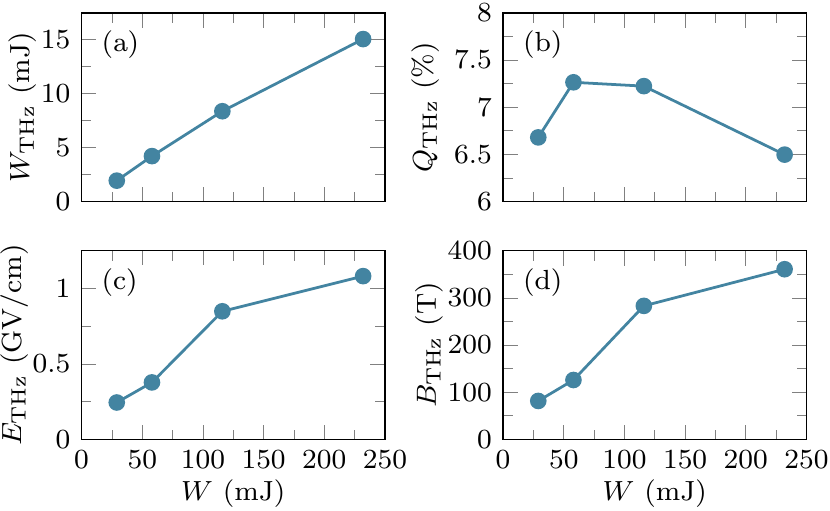}
    \caption{\label{fig:thz_scaling}%
             (a) THz pulse energy $W_\mathrm{THz}$, (b) THz conversion
             efficiency $Q_\mathrm{THz}$, (c) estimated peak electric field
             $E_{\mathrm{THz}}$ of focused THz pulse, and (d) estimated peak
             magnetic field $B_{\mathrm{THz}}$ of focused THz pulse versus input
             energy $W$ of 3.9\,$\mu$m two-color laser pulses.}
\end{figure}

In addition, we studied how the parameters of generated THz radiation scale with
the input energy $W$ of the 3.9\,$\mu$m two-color pulses.
Figure~\ref{fig:thz_scaling}a shows that the energy of the generated THz pulse
growths almost linearly with energy $W$, reaching 15\,mJ at $W=232$\,mJ.
The THz conversion efficiency weakly depends on input energy and is
about 7\,\% (see Fig.~\ref{fig:thz_scaling}b).
The peak THz electric field strengths obtained for $W$=29, 58, 116, and 232\,mJ
are equal to 56, 65, 67, and 77\,MV/cm, respectively.
Considering the almost linear increase of the THz energy with increasing $W$,
and taking into account that the width of the THz spectrum remains the same,
this rather small growth of THz field strength suggests that the spatial size of
the generated THz pulses increases with $W$.
To verify this, we calculated the fluence distributions (intensities integrated
over time) for all THz pulses and estimated their $1/e$ radii $a_\mathrm{THz}$
using a Gaussian fitting.
As a result we obtained $a_\mathrm{THz}$=0.5, 0.77, 1.72, and 2.2\,mm for the
corresponding input energies $W$=29, 58, 116, and 232\,mJ.
Thus, the growth of THz energy with increase of the input laser energy happens
mainly due to increase of the THz beam spatial size.

Although the obtained THz field strengths already exceed the highest values
reported in the literature, one can reach even higher values when focusing the
corresponding THz beams.
As a simple estimate, we recall that for Gaussian beams the field at the focus
of a lens with focal distance $f$ is $L_d/f$ times higher than the initial one.
Here $L_d = 2\pi f_0/c_0a_0^2$ is the diffraction length, where $f_0$ is the
central frequency and $a_0$ is the $1/e$ beam radius.
We apply this estimate for our THz pulses using the previously calculated radii
$a_\mathrm{THz}$ and $f_0$=8\,THz.
This central frequency corresponds to the center of mass of the THz power
spectrum (we found that $f_0$ does not change with $W$).
Figure~\ref{fig:thz_scaling}c shows that the estimated THz field strengths
$E_\mathrm{THz}$ in the focus of a $1^{\prime\prime}$ off-axis parabolic mirror
($f$=25.4\,mm) reach the $GV/cm$ level (1.1\,GV/cm for $W$=232\,mJ).
In turn, the corresponding magnetic fields $B_\mathrm{THz}=E_\mathrm{THz}/c_0$
reach several hundreds of tesla (see Fig.~\ref{fig:thz_scaling}d).

Note that the above estimates are quite relaxed, for the $W$=232\,mJ case, for
example, the above focusing conditions produce a THz focal waist that is about
2.8 times the central THz wavelength.
This means that one could achieve even stronger focusing, like for instance
in~\cite{Shalaby2015}.
This in turn would result in magnetic fields that go beyond the kT regime,
exceeding any laboratory produced quasi-DC magnetic field reported to date by
any means.

{\it Conclusions.}---%
In conclusion, we numerically simulated two-color filamentation of 3.9\,$\mu$m
laser pulses in realistic atmospheric pressure air.
We have shown, that compared to the case of 0.8\,$\mu$m pulses, THz generation
efficiency by 3.9\,$\mu$m two-color pulses is two orders of magnitude higher.
The energy of THz radiation generated by 3.9\,$\mu$m two-color pulses reaches
the multi-millijoule level, the THz electric field strengths can go beyond the
GV/cm level and the magnetic fields can reach the kT.
Such high THz efficiency and energy of THz pulses generated by mid-IR pulses is
the result of several factors, including a novel mechanism where generated high
harmonics contribute to the field symmetry breaking.
The other factors are stronger photocurrents, negligible walk-offs between
harmonics, longer and wider plasma channels.
As a result, we have shown that two-color filamentation of mid-IR laser pulses,
being a source of extremely bright THz radiation, can open the way for future
studies of extreme THz field-matter interactions, nonlinear THz spectroscopy and
imaging.

\begin{acknowledgements}
This work was supported by the National Priorities Research Program grant
No.~NPRP9-383-1-083 from the Qatar National Research Fund (member of The Qatar
Foundation) and the European Union’s Horizon 2020 Laserlab Europe
(EC-GA~654148).
\end{acknowledgements}

%

\section{Supplemental Material}
In our simulations we use the Unidirectional Pulse Propagation Equation
(UPPE)~\cite{Kolesik2002,Kolesik2004,Couairon2011}, given by:
\begin{align} \label{eq:uppe}
    \frac{\partial \hat{E}}{\partial z} = ik_z\hat{E} +
                                          i\frac{\mu_0\mu\omega^2}{2k_z}\hat{N},
\end{align}
where $\hat{E}(k_x,k_y,\omega,z)$ is the spatio-temporal spectrum of the laser
pulse, $\hat{N}(k_x,k_y,\omega,z)$ represents the nonlinear response of the
medium, $k_z(k_x,k_y,\omega)=[k^2(w)-k_x^2-k_y^2]^{1/2}$ is the propagation
constant, $k_x$, $k_y$, and $\omega$ are the spatial and temporal angular
frequencies, $\mu_0$ and $\mu$ are the vacuum and medium permeabilities,
respectively.
The nonlinear response takes into account the third order nonlinear
polarization, $P_{nl}$, the current of free electrons $J_f$ and the current that
is responsible for ionization losses, $J_a$:
\begin{align}
    \hat{N} = \hat{P}_{nl} + \frac{i}{\omega}(\hat{J}_f + \hat{J}_a),
\end{align}
with
\begin{align}
    \hat{P}_{nl} & = \varepsilon_0 \chi^{3} \widehat{E^3}, \\
    \hat{J}_f    & = \frac{q_e^2}{m_e}
                     \frac{\nu_c + i\omega}{\nu_c^2 + \omega^2}
                     \widehat{\rho E}, \\
    \hat{J}_a    & = K\hbar\omega_0
                     \widehat{\frac{\partial \rho}{\partial t} \frac{1}{E}},
                     \label{eq:Ja}
\end{align}
where $\widehat{~}$ denotes the spatio-temporal spectrum, $\varepsilon_0$ is the
vacuum permittivity, $\chi^{3}=4n_0^2\varepsilon_0c_0n_2/3$ is the cubic
susceptibility with $n_2$ being the nonlinear index, $n_0$ is the medium
refractive index at the pulse central frequency $\omega_0$, $c_0$ is the speed
of light in vacuum, $q_e$ and $m_e$ are the charge and mass of the electron,
$\nu_c$ is the collisions frequency, $\rho$ is the concentration of free
electrons (in 1/m$^3$), and $K$ is the order of the multiphoton ionization.
The real part of $\hat{J}_f$ describes inverse Bremsstrahlung, and the imaginary
part is responsible for plasma defocusing.

Together with the UPPE we solve the kinetic equation for plasma
concentration~\cite{Couairon2011}:
\begin{align} \label{eq:kinetic}
    \frac{\partial \rho}{\partial t} = R_1(\rho_{nt} - \rho) + R_2\rho,
\end{align}
where $\rho_{nt}$ is the concentration of neutral molecules, with $R_1$ and
$R_2$ being the optical field and avalanche ionization rates.
To calculate $R_1$ we use the Perelomov-Popov-Terentiev (PPT)
formula~\cite{Perelomov1967}, while $R_2$ is given by
\begin{align}
    R_2 & = \sigma(\omega_0) \frac{E}{U_i},
\end{align}
with
\begin{align*}
    \sigma(\omega_0) = \frac{q_e^2}{m_e}
                       \frac{\nu_c}{\nu_c^2 + \omega_0^2},
\end{align*}
being the inverse Bremsstrahlung cross-section at the pulse central frequency
$\omega_0$, and $U_i$ is the ionization potential.
For the calculation of $\partial \rho/\partial t$ in Eq.~\eqref{eq:Ja} we use
only the first term on the right-hand side of Eq.~\eqref{eq:kinetic}.

\begin{figure}[t]
    \includegraphics{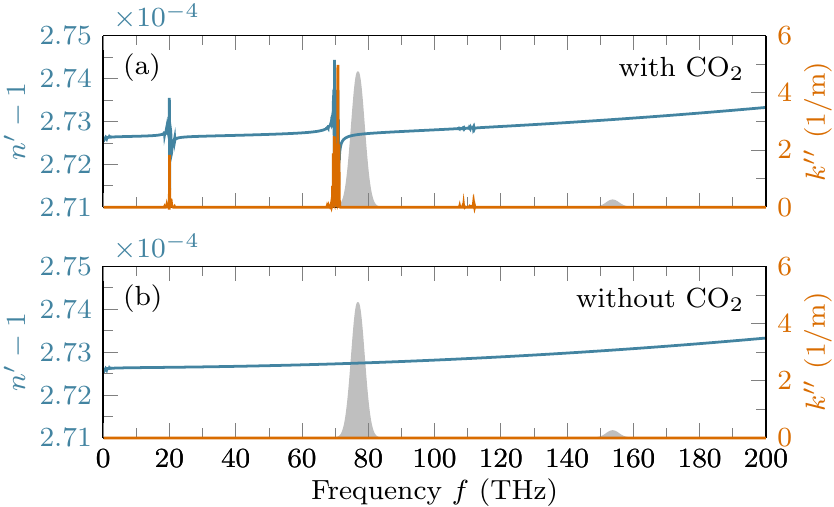}
    \caption{\label{fig:dispersion}%
             The real part $n'$ of the refractive index $n$ (cyan) and the
             absorption coefficient $k''$ (orange) versus frequency $f$ for air
             with (a) and without (b) CO$_2$ gas.
             The spectrum of the initial 3.9\,$\mu$m two-color laser pulse is
             marked by gray.}
\end{figure}

Real atmospheric air is a mixture of several gases, including CO$_2$.
One of the CO$_2$ absorption bands is located at 4.3\,$\mu$m, that is, in close
proximity to the central wavelength of our 3.9\,$\mu$m pulse.
In order to take into account the influence of CO$_2$ gas we use the following
model for complex frequency-dependent refractive index of air
$n=n'+in''$~\cite{Panov2016}:
\begin{equation} \label{eq:n}
    n(\omega) = n_\mathrm{Peck}(\omega) + n_\mathrm{HITRAN}(\omega).
\end{equation}
The real valued refractive index $n_\mathrm{Peck}(\omega)$ is given
in~\cite{Peck1972}, while to calculate the complex refractive index
$n_\mathrm{HITRAN}=n'_\mathrm{HITRAN}+in''_\mathrm{HITRAN}$ we use the data on
spectral lines of atmospheric gases from the HITRAN database~\cite{HITRAN}.
The imaginary part $n''_\mathrm{HITRAN}$ is recalculated from the absorption
coefficient given in the database, then the real part $n'_\mathrm{HITRAN}$ is
restored using the Kramers-Kronig relations.
To study the influence of CO$_2$ resonances we calculated $n_\mathrm{HITRAN}$
for two gas mixtures that represent dry air (i.e., air with zero relative
humidity): the first one with CO$_2$ (0.04\% of CO$_2$, 79.06\% of N$_2$, 20.9\%
of O$_2$) and the second one without CO$_2$ (79.1\% of N$_2$ and 20.9\% of
O$_2$).
The spectral lines for both mixtures are calculated for a temperature of 296\,K
and pressure of 1 atm.
In Fig.~\ref{fig:dispersion} we plot the real part $n'$ of the refractive index
$n$ and absorption coefficient $k''=n''\omega/c_0$ as functions of frequency
$f=\omega/2\pi$ for both gas mixtures.
One can see that the presence of CO$_2$ gas in air gives rise to two absorption
bands centered at 70 and 20\,THz (4.3\,$\mu$m and 15\,$\mu$m, respectively).
However, at this concentration of CO$_2$, the highest values of the absorption
coefficient $k''$ do not exceed several inverse meters.
Therefore for our pulses focused by 200\,mm lens we do not expect such strong
influence of the linear absorption like in~\cite{Panov2016}.
Nevertheless, the presence of CO$_2$ affects the sign of the second order
dispersion coefficient $k_2$ at $\lambda_0$=3.9\,$\mu$m: for the gas mixture
with CO$_2$ the $k_2=-8.61\times10^{-29}$\,s$^2$/m is negative and the
dispersion is anomalous, while for the gas mixture without CO$_2$ the
$k_2=4.18\times10^{-30}$\,s$^2$/m is positive and the dispersion in normal.

In our simulations we assume that the nonlinear index $n_2=10^{-23}$\,m$^2$/W is
the same for $\lambda_0=0.8$ and 3.9\,$\mu$m (the corresponding values of
critical power $P_{cr}$ are 9.65 and 230\,GW); the concentration of neutral
molecules $\rho_{nt}=2.5\times10^{25}$\,1/m$^3$, and collision frequency
$\nu_c=5\times10^{12}$\,1/s.
To calculate the concentration of free electrons, $\rho$, we assumed that air
consists by 79.1\% of N$_2$ and 20.9\% of O$_2$ molecules with ionization
potentials equal to 15.576 and 12.063\,eV, respectively.
For each molecule we solved a separate kinetic equation.

We solved Eq.~\eqref{eq:uppe} on an axially symmetric grid with the following
parameters: the grid size and number of points in spatial domain are 10\,mm and
1000, respectively (spatial resolution is 10\,$\mu$m); the grid size and number
of points in time domain are 10\,ps and 65536, respectively (temporal and
spectral resolutions are 0.15\,fs and 0.1\,THz).

\end{document}